\documentclass[12pt]{iopart}
\usepackage{graphicx}
\usepackage{dcolumn}
\usepackage{bm}

\begin{document}

\title{Effect of Sr substitution on superconductivity in
Hg$_{2}$(Ba$_{1-y}$Sr$_{y}$)$_{2}$YCu$_{2}$O$_{8 - \delta }$
(part2): bond valence sum approach of the hole distribution}

\author{P Toulemonde\dag\footnote[3]{Present address: LPMCN, Universit\'{e} Lyon-I,
B\^{a}timent L\'{e}on Brillouin, 43 Boulevard du 11 Novembre 1918,
F-69622 Villeurbanne cedex, France.}, P Odier\dag, P Bordet\dag, S
Le Floch\dag and E~Suard\ddag}

\address{\dag\ Laboratoire de Cristallographie, CNRS, 25 avenue des martyrs,
BP166, F-38042 Grenoble cedex 09, France.}
\address{\ddag\ Institut Laue-Langevin, 6 rue Jules Horowitz, BP156,
F-38042 Grenoble cedex 09, France.}

\ead{pierre.toulemonde@lpmcn.univ-lyon1.fr}

\date{\today}

\begin{abstract}
The effects of Sr substitution on superconductivity, and more
particulary the changes induced in the hole doping mechanism, were
investigated in
Hg$_{2}$(Ba$_{1-y}$Sr$_{y}$)$_{2}$YCu$_{2}$O$_{8-\delta }$ by a
"bond valence sum" analysis with Sr content from y~=~0.0 to
y~=~1.0. A comparison with CuBa$_{2}$YCu$_{2}$O$_{7-\delta }$ and
Cu$_{2}$Ba$_{2}$YCu$_{2}$O$_{8}$ systems suggests a possible
explanation of the $T_{c}$ enhancement from 0~K for y~=~0.0 to
42~K for y~=~1.0. The charge distribution among atoms of the unit
cell was determined from the refined structure, for y~=~0.0 to
1.0. It shows a charge transfer to the superconducting CuO$_{2}$
plane via two doping channels $\pi(1)$ and $\pi(2)$, i.e. through
O2${_{apical}}$-Cu and Ba/Sr-O1 bonds respectively.

\end{abstract}

\submitto{\JPCM} 

\maketitle

\section{\label{sec:level1}Introduction}

This study is devoted to understand the effects of Sr substitution
in the mercury-bilayer cuprate
Hg$_{2}$Ba$_{2}$YCu$_{2}$O$_{8-\delta }$ (Hg-2212) structure. It
is also often associated to a chemical pressure effect accounting
for the smaller size of Sr substituted to Ba. In part I, we
extracted precise structural data on bond lengths from neutron
powder diffraction ("NPD")~\cite{Toulemonde1}. We studied the
effect of Sr substitution on the barium site in the
Hg$_{2}$(Ba$_{1-y}$Sr$_{y}$)$_{2}$YCu$_{2}$O$_{8-\delta }$ series
and compared to the effect of mechanical pressure. This was
motivated by the previous observation that, as in
(La,Sr)$_{2}$CuO$_{4}$ (LSCO)~\cite{Cava,Torrance,Radaelli1}, Sr
substitution in Hg-2212 enhances $T_{c}$ from 0~K for the
unsubstituted compound to 42~K in the fully substituted
Hg$_{2}$Sr$_{2}$YCu$_{2}$O$_{8-\delta }$ compound. This
enhancement can also be obtained by mechanical pressure which
increases $T_{c}$ of the pure Ba-based Hg-2212 by 50~K under
20~GPa~\cite{Acha}. Moreover, $T_{c max}$ can be pushed to 60~K by
Ca doping in
Hg$_{2}$Sr$_{2}$(Y$_{0.80}$Ca$_{0.20}$)Cu$_{2}$O$_{8-\delta
}$~\cite{Toulemonde2,Toulemonde3}.

In the previous paper~\cite{Toulemonde1}, we showed that Sr
induces an overall shrinkage of the structure due to its smaller
ionic radius than Ba. The detailed analysis of the position shifts
of each atom is summarized as follows. Both Cu and O1 (in-plane)
move up (see figure~5 in ref.\cite{Toulemonde1} for the structure
description and atoms labels), i.e. towards the Ba/Sr site, and
become also closer in their plane. This does not affect the
buckling angle of the superconducting planes (i.e. the Cu-O1-Cu
angle). The Cu-O1 bond length reduction is probably one of the
most important effect contributing to enhance $T_{c}$. Note that
the rate d$T_{c}$/da (where "a" is the unit cell length in
${\AA}$) induced by Sr substitution ($\approx$
850~K.{\AA}${^{-1}}$) is consistent with that observed in
compressed epitaxial films of LSCO ($\approx$
1000~K.{\AA}${^{-1}}$)~\cite{Locquet}. The vertical movement of
O2, Ba/Sr, Hg/Y are modest in comparison to that of O3 that shifts
significantly towards the Hg/Y plane, making it flat. Hence the
"BaO$_{9}$" polyhedron become smaller and better connected to the
superconducting block ("SB", composed of the two CuO$_{2}$
superconducting planes separated by the Y~plane). The "CuO$_{5}$"
pyramids flattens due to the enormous reduction of
Cu-O2$_{apical}$ distance that becomes very small, i.e.
2.27~{\AA}, as compared to unsubstituted Hg-12(n-1)n compounds
where Cu-O2 is close to
2.75-2.80~{\AA}~\cite{Radaelli2,Chmaissem1,Wagner1,Antipov1}. Even
in the fully Sr substituted (Hg,Re)-1212 or
(Hg,Re)-1223~\cite{Chmaissem2} or in the 91 ~{\%} Sr substituted
(Hg,Pb)-1223~\cite{Karpinski1}, Cu-O2 remains larger, being close
to 2.40-2.47~{\AA} in all cases. It is instructive to compare Sr
substitution effect, i.e. the chemical pressure effect, to the
mechanical pressure effect, but this is a complex matter because
chemical pressure does not simply mimic mechanical pressure. In
both cases, the charge reservoir ("CR", composed by the double
(Hg/Y)O3 layer, linked to its two neighboring Ba/SrO2 planes) is
compressed. But, the difference concerns the following: in
Hg-2212, the SB thickness decreases with mechanical pressure
(Hg-2212 data from~ref.\cite{Loureiro,Bordet1}) while it increases
with the chemical pressure~\cite{Toulemonde1}. The same difference
for the SB is also observed between mechanical pressure in
Hg-1223~\cite{Armstrong} and chemical pressure in Sr substituted
CuBa$_{2}$YCu$_{2}$O$_{7-\delta }$ (Cu-1212)~\cite{Licci},
Cu$_{2}$Ba$_{2}$YCu$_{2}$O$_{8}$ (Cu-2212)~\cite{Karpinski2} and
Hg-12(n-1)n layered cuprates~\cite{Chmaissem2,Marezio1}. It cannot
be directly linked with the variation of $T_{c}$ since in Hg-2212
Sr enhances $T_{c}$, while in Cu-1212, Cu-2212 and Hg-12(n-1)n it
decreases $T_{c}$~\cite{Wada,Ganguli,Subramian,Sin}.

It is well known that in layered cuprates, the $T_{c}$ is governed
by the hole concentration in the CuO$_{2}$ planes~\cite{Tallon1}.
However being a mixed valency (in "chemical terms") compound, the
holes are distributed over the Cu (3d) and oxygen site (2p) of the
CuO$_{5}$ polyhedron and in the 3d Cu orbitals (3d$_{z^{2}}$,
3d$_{x^{2}-y^{2}}$). Many efforts have been experimentally done to
find strategies to enhance $T_{c}$ over the last decades and a few
years ago, Brown~\cite{Brown1}, Tallon~\cite{Tallon1} and
Karppinen et al.~\cite{Karppinen1} introduced and used the concept
of "bond valence sum" (BVS), to quantify the relationship between
the atomic structure and electronic properties. It aims to
describe how are distributed charges among the different parts of
the structure to have a better insight in the efficient parameters
promoting $T_{c}$. In this aim, they took benefit of the very
detailed and precise data of bonds lengths extracted from NPD.

In this paper, we will use the refined structure determined from
NPD for three compositions of the
Hg$_{2}$(Ba$_{1-y}$Sr$_{y}$)$_{2}$YCu$_{2}$O$_{8-\delta }$ series:
y~=~0.0, 0.5 and 1.0 (for more details, see tables~1 and~2 in
part~1 of this work in ref.~\cite{Toulemonde1}). The data for the
unsubstituted compound y~=~0 is based on the original Hg-2212
structure determined by Radaelli et
al.~\cite{Radaelli3,Radaelli4}.

The calculation of the formal valence of copper from the refined
compositions (O3 and Y/Hg occupancies), gives 2.13 (assuming
13~{\%} of Y on the Hg site and n(O3)~=~0.88), 2.21 and 2.06 for
y~=~0.0, 0.5 and 1.0 respectively. Then, no coherent correlation
between this Cu valence (or the oxygen content) and the continuous
increase of T$_{c}$ can be done. However, as shown by Alonso et
al.~\cite{Alonso1}, a simple formal valence analysis is not
appropriate in the case of Hg-2212. The doping level of the
CuO${_2}$ planes should be lower than expected from ionic
considerations. As shown in the following discussion, the BVS
approach allows to better estimate the charge distribution into
the 2212 lattice and to identify the different doping channels
involved in the charge transfer induced by Sr substitution.

\section{\label{sec:level2}The "Bond Valence Sum" method}

\subsection{Background}

The BVS method~\cite{Brown1} expresses the charge distribution and
the crystalline stress on the different sites of a crystal
structure.

The BVS are calculated according to equation~1:
\[
Vi = \pm \sum {S_{ij} = \pm \sum {\exp \left[ {\left( {R_{ij} ^0 -
r_{ij} } \right) / B} \right]}~~Eq.(1)}
\]

\noindent where V$_{i}$ is the valence of the ion i, $R_{ij}^0
$(in~{\AA}) is an empirical distance, specific of each i-j pair of
ions (tabulated values can be found in the
literature~\cite{Brown1,Brown2}), r$_{ij}$ is the experimental
bond length (in~{\AA}) of the i-j pair considering only the first
nearest neighbors, and B is an empirical constant equal to 0.37.
The $R_{ij}^0$ values are listed in table~1.

For each site, the calculations give a value which is the sum of
the valence of the site and its stress state (compression or
extension of the coordination polyhedron around the site) with
respect to compounds where the structure is undistorted. The more
stressed site, the more deviation to the valence of the
undistorted structure. The differences between the calculated
V$_{i}$ and the formal valences are then due to the stress induced
by the neighboring sites for atoms having a fixed valence (for
example Ba$^{2 + }$, Sr$^{2 + }$ or Y$^{3 + })$, and/or by a
charge transfer from the neighboring sites for atoms being able to
have an intermediate valence state, for instance copper in
pyramidal coordination (Cu$^{2+\delta}$). This concept initially
developed in ionic compounds has been applied successfully in
strongly correlated systems where electrons are rather localized.
It was used in many compounds, including high-$T_{c}$
superconductors where correlation of the in-plane hole density has
been discussed with $T_{c}$~\cite{Tallon1}. The enhancement of
$T_{c}$ observed in our Hg-2212 series could be correlated with a
modification of the charge transfer which occurs between the CR
and the SB. In such a case, a BVS analysis could quantify this
modification. The values of the Cu-O2$_{apical}$ bond length
(around~2.27~{\AA} for y~=~1.0) and the buckling angle of the
CuO${_2}$ superconducting planes (around 14~deg.) are close to
those measured in Cu-1212 (one Cu-O chain in the CR) and Cu-2212
(double Cu-O chain in the CR) compounds, then similar changes of
the charge distribution can be expected as in the Sr substituted
Hg-2212, Cu-1212 and Cu-2212 systems. Then, our BVS analysis will
compare Sr substituted Cu-1212~\cite{Licci},
Cu-2212~\cite{Karpinski2} and also (Hg,Pb)-1223 (Ca instead of Y
between the CuO${_2}$ planes)~\cite{Karpinski1} with our Sr
substituted Hg-2212.

\subsection{BVS analysis of Hg$_{2}$(Ba$_{1-y}$Sr$_{y}$)$_{2}$YCu$_{2}$O$_{8-\delta }$}

In our BVS calculations a modified 2212 structure of the y~=~0
composition (initially proposed by
Radaelli~et~al.~\cite{Radaelli3,Radaelli4}), was used in order to
be consistent with our refinements, i.e. Hg is partly substituted
by Y, as shown by S.M.~Loureiro~\cite{Loureiro} and confirmed by
our previous work~\cite{Toulemonde2}. We kept the atomic positions
determined by Radaelli et al. and replaced the Y/Hg and O3
occupancies with the values refined by S.M. Loureiro: 0.13/0.87
and 0.88 (instead of 0.78) respectively. This choice will affect
only the BVS of the atoms bonded to O3 or Hg/Y, i.e. Hg/Y and O3
themselves and not significantly Ba whose BVS is more affected by
the four Ba-O1 and four Ba-O2 contributing bonds.

One can have two different approaches to calculate the
contribution of sites which are occupied by two kind of cations:
Hg$^{2 + }$/Y$^{3 + }$, Ba$^{2 + }$/Sr$^{2 + }$ or
Cu$^{2+\delta}$. The first one is to consider each atom
individually with its specific $R_{ij}^0$ and the BVS of the site
is calculated from the weighted sum of the two individual BVS. In
this calculation we consider in first approximation a unique
coordination number for both cations. For Ba/Sr sites and y~=~0.5
for example, one will add a contribution calculated considering a
site with 0.5 Ba (characterized by $R_{ij}^0 (Ba^{2 + } - O) =
2.285\;\mbox{{\AA}})$, to a second one with 0.5~Sr ($R_{ij}^0
(Sr^{2 + } - O) = 2.118\;\mbox{{\AA}})$. That was the point of
view adopted by Karpinski et al. in the calculation of the copper
BVS in the two Cu sites of Cu-2212~\cite{Karpinski2}. The second
approach considers that the cation-oxygen distances calculated
from the refinement of NPD data represent a spatial average of the
sites which already takes into account the substitution effects.
In this case, for each bond, one use an average $R_{ij}^0$ value
and the corresponding refined $r_{ij}$ distance which is in fact
already averaged. For instance, to calculate the BVS of a Ba/Sr
site, a fully occupied site with an average $R_{ij}^0$ of Ba-O and
Sr-O pairs is considered. This was the method used by Licci et al.
in the compound Cu-1212~\cite{Licci}. Here, we used both
approaches, and because the differences between $R_{ij}^0 $ are
not large, both give BVS values which are very close to each
others (within a 2~{\%}). Consequently, both approaches give the
same tendencies versus the Sr content. We present here only the
BVS values calculated by the second method.

To summarize (see table1), the value $R_{ij}^0(Cu-O)$ of the Cu
site was calculated by assuming a linear dependence of $R_{ij}^0$
between the limiting values, i.e. 1.679~{\AA} for $R_{ij}^0 (Cu^{2
+ } - O)$ and 1.73~{\AA} for $R_{ij}^0 (Cu^{3 + } - O)$, on the
basis of the formal valence calculated from the NPD refined
composition. It gives: $R_{ij}^0(Cu-O)$~=~1.686, 1.690 and
1.682~{\AA} for y~=~0.0, 0.5 and 1.0 respectively. We note here
that $R_{ij}^0(Cu-O)$ is not changed among the series. Then, the
variation of the BVS of Cu is essentially caused by the shrinkage
of the Cu-O1 (i.e. the a-axis) and Cu-O2 distances, and not by the
insignificant change of its $R_{ij}^0$. The O3 occupation factor
was taken into account for the BVS calculations of Ba/Sr and Hg/Y
sites. For the Ba/Sr site, the weighted (nominal Sr content)
average of $R_{ij}^0 (Ba^{2 + } - O)~=~2.285\;\mbox{{\AA}}$ and
$R_{ij}^0 (Sr^{2 + } - O) = 2.118\;\mbox{{\AA}}$ was taken into
account. For the Hg/Y site, we took the weighted (i.e. the refined
Y content: 13, 17 and 24~{\%}~for y~=~0.0, 0.5 and 1.0
respectively) average of $R_{ij}^0 (Hg^{2 + } - O)$~=~1.972~{\AA}
and $R_{ij}^0 (Y^{3 + } - O)$~=~2.014~{\AA}. In the BVS
calculation of the Hg/Y site, Hg/Y-O2, Hg/Y-O3 (vertical) and the
four different planar Hg/Y-O3 bond distances were considered. The
whole BVS calculations for each site were done considering the
first neighbors. All the values are summed up in table~1. In most
published calculations, the BVS error bar is often not indicated,
or calculated, or is underestimated. In our case, the errors were
estimated by taking into account not only the incertitude of the
refined distances but also the error bars of the refined occupancy
factors (of O3 for example) and the $R_{ij}^0$ error bars issued
from the precision of the refined composition. In all cases, the
different changes observed are higher than the error bars.
Although our BVS values are rather close to the ionic formal
valences values, some differences remain, expressing charges
transfer that is now discussed.

\begin{table}
\caption{\label{tab:table3}Bond Valence Sums (BVS) of cation and
oxygen sites in Sr substituted Hg-2212 calculated from the
Rietveld refined structure. The $R_{ij}^0$ constants are based on
arithmetic mean values weighted by the the refined composition.
For Cu-O bond, tabulated $R_{ij}^0$(Cu$^{3 + }$-O) and
$R_{ij}^0$(Cu$^{2 + }$-O) values were considered, weighted to fit
the copper valence determined from refined composition.}
\begin{indented}
\item[]\begin{tabular}{@{}ccccccccc}
\br
 Sr content& & Hg/Y& Ba/Sr& Y& Cu& O1& O2& O3
\\ \hline \raisebox{-1.50ex}[0cm][0cm]{0.0}& $R_{ij}^0$ &
$1.977$& $2.285$& $2.014$& $1.686$& -& -& - \\
 &
BVS& 2.04(4)& 2.12(2)& 2.80(1)& 2.13(2)& -2.07(1)& -2.01(3)&
-1.75(3) \\ \hline \raisebox{-1.50ex}[0cm][0cm]{0.5}& $R_{ij}^0$ &
$1.979$& $2.202$& $2.014$& $1.690$& -& -& - \\
 &
BVS& 2.21(4)& 1.92(4)& 2.85(1)& 2.27(2)& -2.07(1)& -1.95(2)&
-1.86(3) \\ \hline \raisebox{-1.50ex}[0cm][0cm]{1.0}& $R_{ij}^0 $
&
$1.982$& $2.118$& $2.014$& $1.682$& -& -& - \\
 &
BVS& 2.11(4)& 1.72(2)& 2.83(1)& 2.32(2)& -2.06(1)& -1.84(2)&
-1.79(3) \\ \br
\end{tabular}
\end{indented}
\end{table}

\subsection{Charges transfer in cuprates}

The carrier concentration in cuprates controls directly their
superconducting properties. Its determination and distribution
over the sites of the unit cell is then very important, but not
trivial. It may be modified by oxygen doping or charge
compensation after heterovalent substitution. One must distinguish
compounds in which oxygen dope the Cu-O chains (in Cu-1212 for
example) from those where the doping oxygen is located in the
center of the Hg/Tl squares (Hg-12(n-1)n, Tl-12(n-1)n or
Tl-22(n-1)n families). In addition to different experimental
methods (thermogravimetric, wet-chemical redox analysis, XANES or
XPS spectroscopies), the BVS method based on a bond-valence-length
correlation is very convenient for identifying the charges of the
different sites. As shown by Karppinen~et~al.~\cite{Karppinen1},
one must distinguish three different hole-doping channels to the
superconducting planes: $\pi(1)$ through a shortening of the Cu-O2
(i.e. apical oxygen) bond, $\pi(2)$ through a lengthening the
O1-(Ba/Sr) bond (i.e. in-plane oxygen) or $\pi(3)$ through a
lengthening the O1-(Y/Ca) bond.

When oxygen dope the charge reservoir, channels $\pi(1)$ and
$\pi(2)$ are activated in the case of Cu-1212, while only the
channel $\pi(1)$ is activated in the case of Hg-1201 or Hg-1212
(and Tl-12(n-1)n or Tl-22(n-1)n families also)~\cite{Karppinen2}.
Generally, channel $\pi(3)$ is activated when the doping concerns
the site located in-between the CuO$_{2}$ planes,
Ca$^{2+}$/Y$^{2+}$ for example. For Sr substitution in Hg-2212,
one expect that the charge transfer occurs principally through
channel $\pi(1)$, i.e. by a charge transfer along the
Cu-O2$_{apical}$ bond and secondly through channel $\pi(2)$, i.e.
along the Ba/Sr-O1 bond, because Cu-O2 bond length decreases by
-8.0~{\%}, O1-(Ba/Sr) by -2.8~{\%} and O1-Y remains unchanged
(-0.3~{\%}), as illustrated in figure~10 of
part~1~\cite{Toulemonde1}.

An other interesting feature, observed in (Y$_{1 -
x}$Ca$_{x}$)Ba$_{2}$Cu$_{3}$O$_{7 - \delta }$ (i.e. Cu-1212) by
Merz~et~al.~\cite{Merz} using X-ray absorption spectroscopy,
concerns $T_{c~max}$ that is not only controlled by the hole
concentration in the CuO$_{2}$ planes, but by the presence of
holes in the apical oxygen site. Even though the CuO$_{2}$ planes
do contain a sufficiently large hole concentration to
superconduct, superconductivity to occur needs the presence of
holes in the apical site. Moreover, raising the oxygen content in
the charge reservoir (i.e. in the Cu-O chains) increases the holes
concentration of both in-plane and apical oxygen sites. In
contrary holes introduced by replacing Y$^{3+}$ by Ca$^{2+}$
appear solely in the CuO$_{2}$ planes and enable (or enhance)
superconductivity only if a minimum of holes were already present
on the apical oxygen site.

\section{Discussion}

After having presented the different channels in charge transfer,
we use BVS data and this scheme to discuss the Sr substitution
effect.

\begin{figure}
\begin{center}
\includegraphics[width=0.6\linewidth]{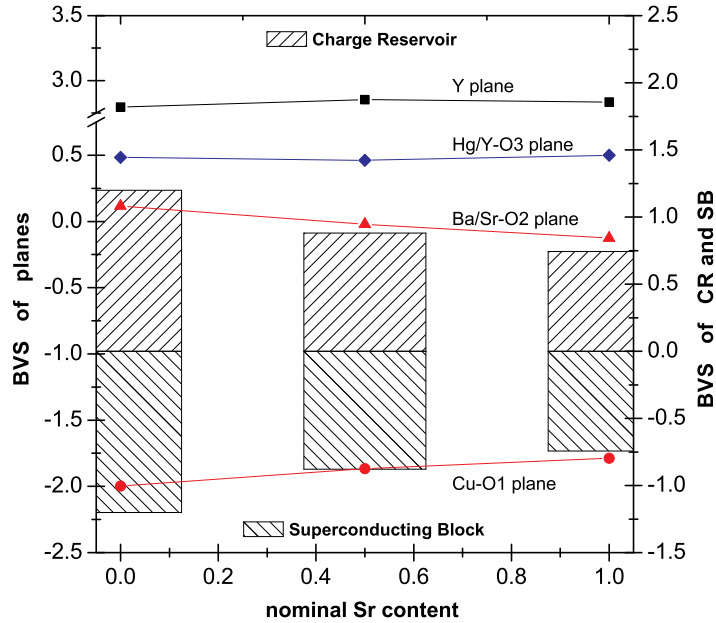}
\end{center}
\caption{\label{fig:eps13}Left scale: BVS variations of the Y,
CuO1, BaO2 and (Hg,Y)O3 planes of Hg-2212 versus Sr content
(points), calculated considering the refined composition. Right
scale: the resulting BVS of the CR and SB (bars) are also plotted.
The error bars are not indicated for more clarity.}
\end{figure}

The figure~1 shows  the changes of BVS versus the Sr content for
the CR and SB (already defined in introduction) and their
constitutive layers. The BVS of CR and SB was calculated by
summing the corresponding BVS values for each site belonging to
the CR and SB blocks. The same was done for the atomic planes and
plotted on the same figure. The very clear trend is an increase of
the BVS for the SB and a symmetrical decrease for the CR. Both
variations show a charge redistribution of their constituting
planes (two CuO1 and one Y planes for SB, and two Hg/YO3 and two
Ba/SrO2 planes for CR). While no change of BVS is observed for Y
and Hg/Y-O3 planes, only BVS of Ba/Sr-O2 and Cu-O1 layers are
affected by the Sr substitution.

\begin{figure}
\begin{center}
\includegraphics[width=0.5\linewidth]{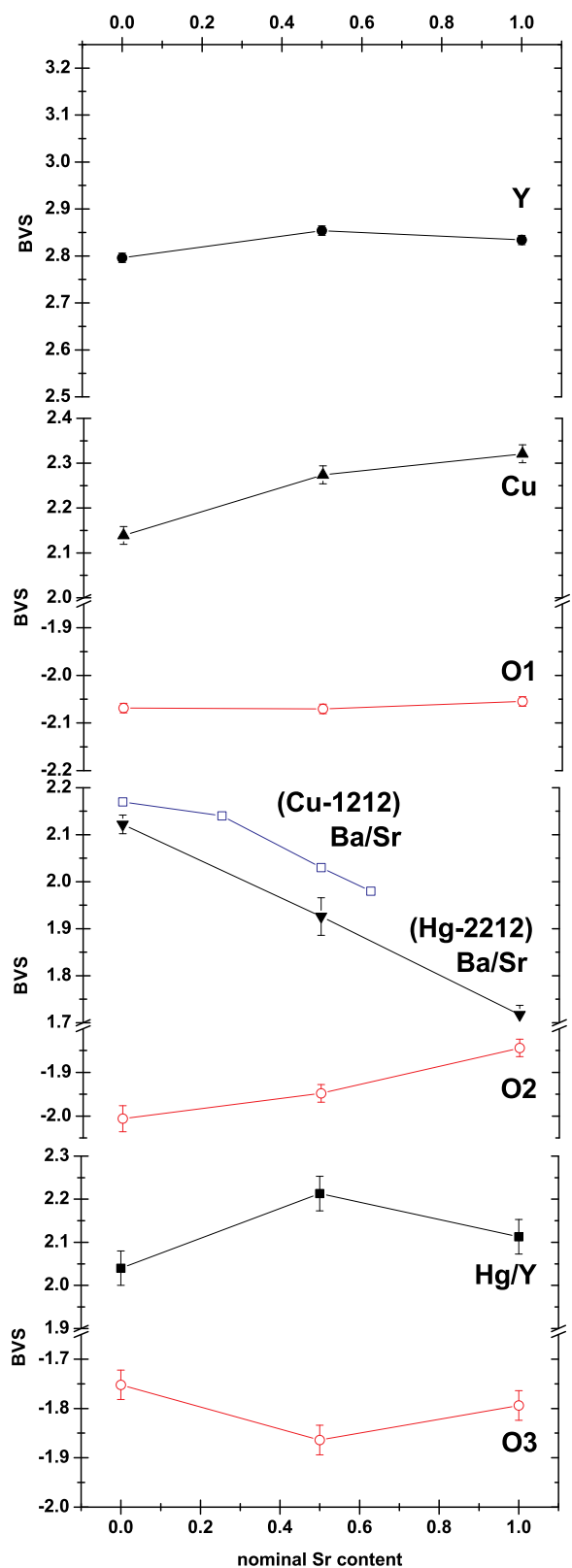}
\end{center}
\caption{\label{fig:eps14}BVS variations of the atoms in Hg-2212
versus Sr content. The BVS variation for the equivalent Ba/Sr site
in Cu-1212~\cite{Licci} is also plotted for comparison.}
\end{figure}

Let us now discuss the effects atom by atom. This is illustrated
in figure~2 showing the BVS variations for each atom. The four
graphs are stacked in the same order as in the 2212 structure and
drawn at the same scale to help a direct comparison. The BVS of Y
($\approx$~2.8) and O1 ($\approx$~-2.06) sites remain nearly
constant, as in Cu-1212~\cite{Licci} or Cu-2212~\cite{Karpinski2}
(not drawn on fig.~2). The main effect of Sr substitution is a
strong decrease of the BVS of the Ba/Sr site, accompanied by an
important increase of the BVS of its neighboring O2 oxygen and of
Cu. This is the signature of a significant charge transfer from
the CR to the SB essentially affecting the superconducting
CuO${_2}$ plane.

\begin{figure}
\begin{center}
\includegraphics[width=0.6\linewidth]{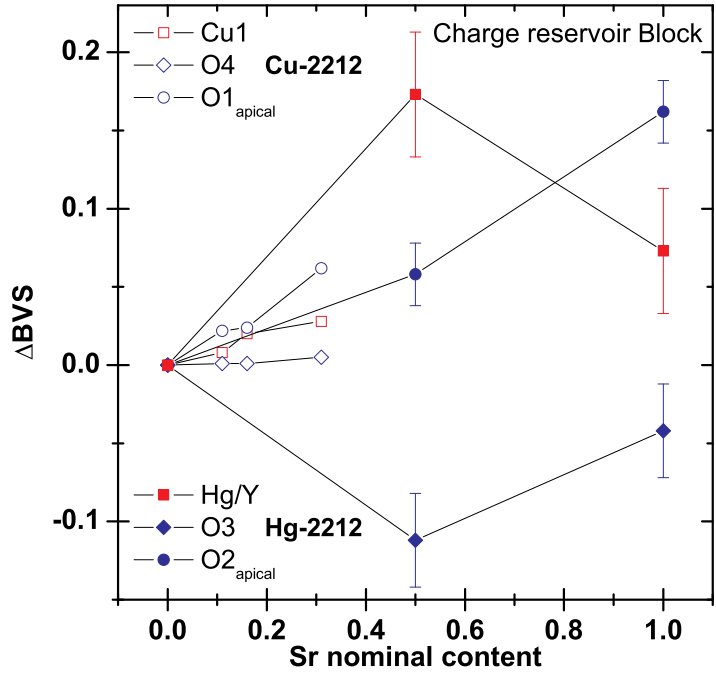}
\end{center}
\caption{\label{fig:eps15}Relative BVS Variations of the cationic
site of the C.R. (i.e. Cu1 for Cu-2212~\cite{Karpinski2}, Hg/Y for
Hg-2212), oxygen of the C.R. (O4 for Cu-2212, O3 for Hg-2212) and
apical oxygen (O1 for Cu-2212 and O2 for Hg-2212) versus Sr
content.}
\end{figure}

The modest variation of the BVS of the Hg/Y site (in the CR) is
related to the increasing substitution of the Y$^{3 + }$ (table~1
in part~1~\cite{Toulemonde1}) compensated by the increasing O3
oxygen content. The BVS of Hg/Y and O3 sites do not change
monotonically with Sr content, they cannot be therefore directly
correlated to the $T_{c}$ increase.

In figure 2 is also plotted the BVS variation of the Ba/Sr site in
Cu-1212~\cite{Licci} compared to that in Hg-2212. The comparison
to the formal value of the Ba/Sr valence (+2) quantifies the
stress of the site. In both cases, for x~$<$~0.5, the large
cationic Ba site in y~=~0.0 composition is compressed
(BVS~-~2~$>$~0). This stress, which plays an important role in the
charge transfer from the CR to the
SB~\cite{Licci,Marezio2,Marezio3,Marezio4}, is reduced by Sr
substitution. The decrease of this stress with Sr substitution is
quite similar in Hg-2212 and Cu-1212, however $T_{c}$ increases in
Hg-2212 while it decreases in Cu-1212. Then, at first view, this
parameter does not seem to be of first order in the enhancement
mechanism of $T_{c}$ in Sr substituted Hg-2212. However, the
redistribution of the charges associated to this stress relaxation
is the important point to consider. One notices also that this
release is correlated with the increase of the thermal Debye
Waller factor from 0.5~{\AA}$^{2}$ to a value higher than
1.5~{\AA}$^{2}$ for y~=~1.0 (see table~1 in
ref.\cite{Toulemonde1}), the Sr-O bond being stretched for this Sr
content.

The comparison of the behavior of Hg/Y site with the equivalent
site in the Cu-2212 compound~\cite{Karpinski2}, i.e. the copper
atom of the Cu1(O1,O4) planar square in the CR, is interesting and
shows a contrasted behavior, figure~3 ($\Delta $BVS i.e. the
variation of BVS with respect to the composition y = 0.0). The
data plotted on this figure are not corrected for stress. When
this is done (as explained in ref.~\cite{Karpinski2}), no
variation of the BVS of O4 (in the chains of Cu1 in Cu-2212) is
observed while a charge transfer occurs from the apical O1 oxygen
(whose BVS decreases) to the Cu1 site (whose BVS increases). For
Hg-2212, such charge transfer from apical O2 oxygen to the
equivalent Hg/Y site does not occur.

\begin{figure}
\begin{center}
\includegraphics[width=0.6\linewidth]{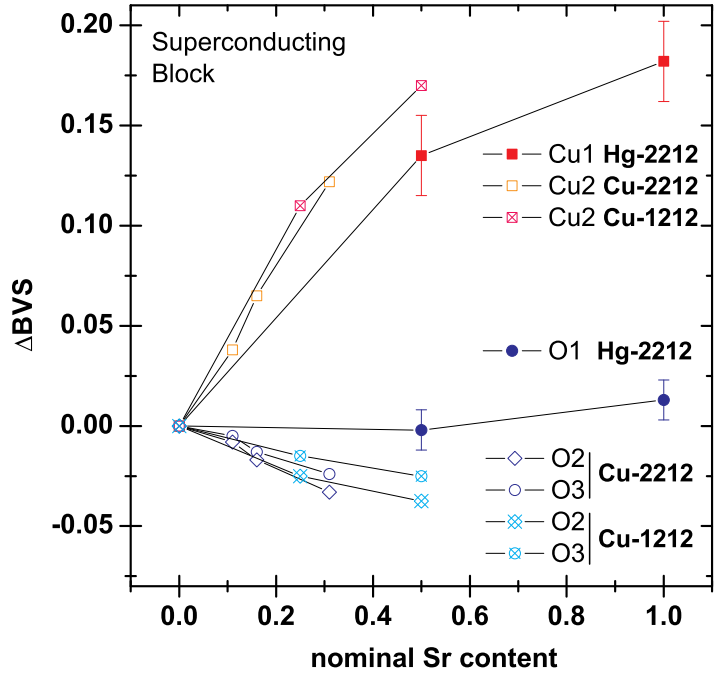}
\end{center}
\caption{\label{fig:eps16}Relative BVS variations of copper and
oxygen of the SB versus Sr content in Cu-1212~\cite{Licci},
Cu-2212~\cite{Karpinski2} and Hg-2212.}
\end{figure}

The figure 4 compares $\Delta $BVS for Cu and O atoms of the
superconducting CuO$_{2}$ planes for Cu-1212~\cite{Licci},
Cu-2212~\cite{Karpinski2} and Hg-2212 (this work). When Sr is
introduced, the BVS of Cu increases for those three compounds. In
Cu-1212 and Cu-2212 the BVS of oxygen atoms (i.e.~O2~and O3 in
Cu-1212 and Cu-2212) decreases symmetrically, as expected for
in-plane charge transfers~\cite{Licci,Karpinski2}. Surprisingly,
the BVS for the in-plane O1 oxygen in Hg-2212 behaves differently,
staying nearly constant whatever Sr is. This indicates a different
mode of charge transfer between Hg-2212 and Cu-1212 or Cu-2212
cuprates. We assume that the stress relaxation of the Ba/Sr site,
i.e. its movement towards O3, occurs differently and opens
different channels of charge transfer (corresponding to different
effect on Cu-O2, Ba/Sr-O1 and Y-O1 bonds).

The figure~5 shows for Hg-2212 the net hole concentration of the
equivalent CuO${_2}$ planes
"p(CuO${_2}$)"~\cite{Tallon1,Karppinen1} versus the nominal Sr
content. It is defined as p(CuO${_2}$)~=~V(Cu) + 2V(O1) + 2, i.e.
equal in fact to the sum of $\pi(1) + \pi(2) + \pi(3)$, $\pi(1)$,
$\pi(2)$ and $\pi(3)$ being respectively the three different
contributions associated to the three channels:
$\pi(1)~=~S{_{Cu-O2}}$, $\pi(2)~=~4S{_{Ba/Sr-O1}}$ and
$\pi(3)~=~4S{_{YO1}}$ (see paragraph 2.3 and Eq.(1)). These three
partial hole concentrations to the total value p(CuO${_2}$) are
also shown in the right scale of figure~5. The contributions of
channels $\pi(1)$ and $\pi(2)$ increase with the Sr content, while
the contribution of channel $\pi(3)$ remains negative and
constant. In other words, the relaxation of the stress on the Ba
site by Sr substitution activates the channels $\pi(1)$ and
$\pi(2)$, i.e. the charge transfer from the CR to the SB via
O2${_{apical}}$-Cu and Ba/Sr-O1 bonds. A similar evolution is also
observed in the monolayer (Hg,Pb)-1223 substituted with
Sr~\cite{Karpinski1}. Moreover, in this 1223 compound the $\pi(3)$
contribution increases with Sr content, while in Hg-2212 it
remains constant and negative ($\approx$~-0.4). Indeed, for
Hg-2212 no significant change of Y-O1 bond occurs when Sr is
substituted to Ba, in comparison to (Hg,Pb)-1223 where the
equivalent Ca-O1 bond increases by 0.02~{\AA}~\cite{Karpinski1}.
This is probably a major difference that could explain explain why
$T_{c}$ decreases when Sr is substituted to Ba in (Hg,Pb)-1223
while it increases in Hg-2212.

\begin{figure}
\begin{center}
\includegraphics[width=0.6\linewidth]{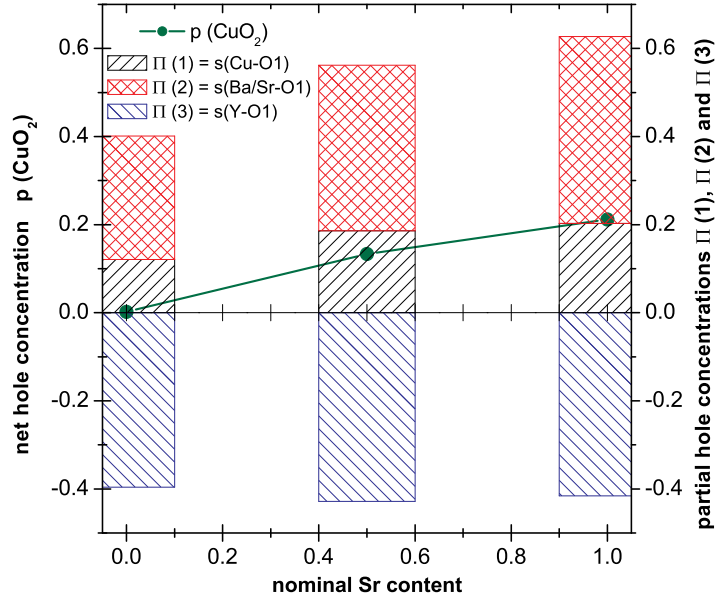}
\end{center}
\caption{\label{fig:eps17}Effect of Sr content on the net hole
concentration p(CuO$_{2}$) in the CuO$_{2}$ planes of the Hg-2212
series (points, left scale). The three different contributions
$\pi(1)$, $\pi(2)$ and $\pi(3)$ to p(CuO$_{2}$) are also plotted
(bars, right scale). The error bars are not indicated for more
clarity.}
\end{figure}

In conclusion, the BVS calculations show how Sr substitution
impact on the charge distribution in the Hg-2212 layered
structure. Two layers play a major role in this charge
redistribution: Ba/SrO2 and CuO1 sheets, while Hg/YO3 and Y sheets
are not important in this respect. The release of stress on the
Ba/Sr site by Sr substitution induces a charge transfer from the
Ba/SrO2 layer (and not from the Y plane) to the CuO${_2}$ plane.
This model may be the key for explaining the increase of $T_{c}$
with Sr content. In addition, it would be interesting to compare
the charge distribution of Hg-2212 compressed by Sr substitution
with the one obtained from mechanical compression, but this kind
of study requires to know very precisely the Hg-2212 structure
under high pressure (up to 7-10~GPa), and particularly the oxygen
positions. Unfortunately, such data (neutron powder diffraction)
are rare~\cite{Loureiro,Bordet1} and not available in the
literature for Hg-2212. Theoretical electronic structure of
Hg-2212 would be also very useful; nevertheless, this calculation
was only made recently for the Hg$_{2}$Ba$_{2}$YCu$_{2}$O$_{8}$
composition~\cite{Alonso1}.

\section{\label{sec:level5}Conclusion}

Following the first part of this work, devoted to the study of
structural effects induced by the Sr substitution in
Hg$_{2}$(Ba$_{1-y}$Sr$_{y})_{2}$YCu$_{2}$O$_{8-\delta}$ (by
neutron powder diffraction) and its effects on $T_{c}$
(enhancement from 0~K for y~=~0.0 to 42~K for y~=~1.0), this
second part is focused on an analysis of the charge distribution
into the 2212 lattice using the refined structures obtained
previously and BVS calculations.

The bond valence sum approach gives a relevant idea of the doping
mechanism related to the Sr substitution. From the BVS analysis,
we demonstrate that the shrinkage of the Cu-O1 bond of the
superconducting plane is accompanied by a reduction of the stress
on the Ba/Sr site which favors a charge transfer from the Ba/Sr-O2
plane to the CuO$_{2}$ superconducting plane by activating the two
doping channels $\pi(1)$ and $\pi(2)$, i.e. through
O2${_{apical}}$-Cu and Ba/Sr-O1 bonds respectively and not via
$\pi(3)$, i.e. through Y-O1 bonds. This could explain the observed
$T_{c}$ increase with the Sr content.

\ack P. Toulemonde thanks CNRS for its financial support during
its PhD research.


\section*{References}
\begin{thebibliography}{42}
\bibitem{Toulemonde1} P. Toulemonde, P. Odier, P. Bordet, S.
Le~Floch, and E.~Suard, submitted to J.~Phys.: Condens.~Matter.
\bibitem{Cava} R. J. Cava, R. B. van Dover, B. Batlogg, and E. A. Rietman,
Phys. Rev. Lett. {\bf 58}, 408 (1987).
\bibitem{Torrance} J. B. Torrance, Y. Tokura, A. I. Nazzal, A. Bezinge, T.~C.~Huang, and S. S. P. Parkin, Phys. Rev. Lett. {\bf 61}, 1127 (1988)
\bibitem{Radaelli1} P. G. Radaelli, D. G. Hinks, A. W. Mitchell, B. A. Hunter, J.
L. Wagner, B. Dabrowski, K. G. Vandervoort, H. K. Viswanathan, and
J. D. Jorgensen, Phys. Rev. B {\bf 49}, 4163 (1994).
\bibitem{Acha} C. Acha, S. M. Loureiro, C. Chaillout, J. L. Tholence, J. J.
Capponi, M. Marezio, and M.~Nunez-Regueiro, Solid State
Communications {\bf 102}, 1 (1997).
\bibitem{Toulemonde2} P.~Toulemonde, PhD thesis, 2000, Grenoble, France.
\bibitem{Toulemonde3} P. Toulemonde and P. Odier, Physica C {\bf 402}, 152 (2004).
\bibitem{Locquet} J.-P. Locquet, J. Perret, J. Fompeyrine, E. M\"{a}chler, J.
W. Seo, and G. Van Tendeloo, Nature {\bf 394}, 453 (1998).
\bibitem{Radaelli2} P. G. Radaelli, J. L. Wagner, B. A. Hunter, M. A. Beno,
G. S. Knapp, J. D. Jorgensen, and D. G. Hinks, Physica~C~{\bf
216}, 29 (1993).
\bibitem{Chmaissem1} O. Chmaissem, Q. Huang Q, E. V. Antipov, S. N. Putilin,
M. Marezio, S. M. Loureiro, J. J. Capponi, J. L. Tholence, and A.
Santoro, Physica~C~{\bf 217}, 265 (1993).
\bibitem{Wagner1} J. L. Wagner, B. A. Hunter, D. G. Hinks, and J. D.
Jorgensen, Phys. Rev. {\bf 51}, 15407 (1995).
\bibitem{Antipov1} E.~V. Antipov, A.~M. Abakumov, and S.N.~Putilin, Supercond. Sci. Technol. {\bf15}, R31 (2002).
\bibitem{Chmaissem2} O. Chmaissem, J. D. Jorgensen, K. Yamaura, Z. Hiro\"{\i},
M. Takano, J. Shimoyama, and K. Kishio, Phys. Rev.~B {\bf 53},
14647 (1996).
\bibitem{Karpinski1} J. Karpinski, H. Schwer, S. M. Kazakov, M. Angst, J. Jun, A. Wisniewksi and R.~Puzniak, Physica~C {\bf 341-348},
421 (2000).
\bibitem{Loureiro} S. M. Loureiro, PhD thesis, 1997, Grenoble, France.
\bibitem{Bordet1} P. Bordet, S. M. Loureiro, J. J. Capponi, and P.~G.~Radaelli,
17th European Crystallographic Meeting (1997).
\bibitem{Armstrong} A. R. Armstrong, W. I. F. David, I. Gameson, P. P. Edwards,
J. J. Capponi, P. Bordet, and M. Marezio, Phys. Rev. B {\bf 52},
15551 (1995).
\bibitem{Licci} F. Licci, A. Gauzzi, M. Marezio, P. Radaelli, R. Masini, and
C. Chaillout-Bougerol, Physical Review B {\bf 58}, 15208 (1998).
\bibitem{Karpinski2} J. Karpinski, S. M. Kazakov, M. Angst, A. Mironov, M.~Mali,
and J. Roos, Phys. Rev. B {\bf 64}, 094518 (2001).
\bibitem{Marezio1} M. Marezio and F. Licci, Physica C {\bf 282-287}, 53 (1997).
\bibitem{Wada} T. Wada, S. Adachi, T. Mihara, and R. Inaba, Jpn. J. Appl.
Phys., Part 2 {\bf26}, L706 (1987).
\bibitem{Ganguli} A. K. Ganguli, M. A. Subramanian, J. Solid State
Chem. {\bf90}, 382 (1991).
\bibitem{Subramian} M. A. Subramian, and M. H. Whangbo, J. Solid State Chem. {\bf 109},
410 (1993).
\bibitem{Sin} A. Sin, F. Alsina, N. Mestres, A. Sulpice, P. Odier, and M.
N\'{u}\~{n}ez-Regureiro, J. Solid State Chem. {\bf 161}, 355
(2001).
\bibitem{Tallon1} J.~L. Tallon, Physica~C~{\bf 168}, 85 (1990).
\bibitem{Brown1} I. D. Brown and D. Altermatt, Acta Crystallog., Sect. B.:
Struct. Sci.  {\bf B41}, 244 (1985).
\bibitem{Karppinen1} M. Karppinen and H. Yamauchi, Mat. Sci. and
Engeneering, {\bf 26}, 51 (1999) and references herein.
\bibitem{Radaelli3} P. G. Radaelli, M. Marezio, M. Perroux, S. de Brion, J.L.
Tholence, Q.~Huang, and A. Santoro, Science {\bf 265}, 380 (1994).
\bibitem{Radaelli4} P. G. Radaelli, M. Marezio, J. L. Tholence, S. d. Brion, A.
Santoro, Q.~Huang, J. J. Capponi, C. Chaillout, T. Krekels, and G.
V. Tendeloo, Journal of Physics and Chemistry of Solids {\bf 56},
1471 (1995).
\bibitem{Alonso1} R.~E.~Alonso, C.~O. Rodriguez, and N.E. Christensen, Phys. Rev. B {\bf 63}, 134506 (2001).
\bibitem{Brown2} I. D. Brown, J. Solid State Chem. {\bf 82}, 122
(1989); {\bf 90}, 155 (1991).
\bibitem{Karppinen2} M. Karppinen and H. Yamauchi, Int. J. Inorg. Mater. {\bf 2}, 589 (2000).
\bibitem{Merz} M. Merz, N. Nücker, P. Schweiss, S. Schuppler, C. T. Chen,
V. Chakarian, J. Freeland, Y. U. Idzerda, M. Kläser, G.
Müller-Vogt, and Th. Wolf, Phys. Rev. Lett. {\bf 80}, 5192 (1998).
\bibitem{Marezio2} M. Marezio, and F. Licci, Supercond. Sci. Technol. {\bf13}, 451 (2000).
\bibitem{Marezio3} M. Marezio, F. Licci, and A.~Gauzzi, Physica~C {\bf337}, 195 (2000).
\bibitem{Marezio4} M. Marezio, E.~Gilioli, P.G.~Radaelli, and F.~Licci, Physica~C {\bf341-348}, 375 (2000).

\end{thebibliography}
\end{document}